\documentstyle[pra,aps,epsf,floats]{revtex}
\begin{document}
\draft

\twocolumn[\hsize\textwidth\columnwidth\hsize\csname @twocolumnfalse\endcsname
\title{\hfill {\small ITP Preprint Number NSF-ITP-96-53}  \\
\vspace{10pt}
Properties of Paired Quantum Hall States at $\nu=2$}
\author{F. G. Pikus\ $^a$ and A. M. Tikofsky\ $^b$}
\address{\ $^a$ Department of Physics, University of California,
Santa Barbara, CA 93106}
\address{\ $^b$ Institute for Theoretical Physics, University of California,
Santa Barbara, CA 93106}
\date{\today}
\maketitle
\begin{abstract}
The energetic properties of a paired quantum Hall state at 
Landau level filling $\nu=2$ are investigated using
variational Monte Carlo techniques.
Pairing is found to be energetically favorable in small magnetic fields
because it introduces correlations between up and down spins that are
absent in the traditional $\nu=2$ state.
We find that pairing survives extrapolation to the thermodynamic limit.
\end{abstract}
]
\narrowtext

The traditional view of the Integral Quantum Hall
Effect (IQHE) treats the system as a gas of non-interacting
electrons in a background magnetic field.
The corresponding spectrum consists of Landau levels 
separated from each 
other by an energy gap $\hbar\omega_c=\hbar eB/m^*c$
where $m^*$ is the band effective mass of the electrons and $B$ is the
external magnetic field. 
It is presumed that $B$ is so large that
many-particle interactions can essentially be ignored because they 
encourage the occupation of higher Landau levels.  Recent 
experiments \cite{Jiang1,Jiang2,Pepper,Tsui}
 have led one of the authors \cite{tikof}
to propose that the traditional picture is incomplete.
For small enough $B$, it was argued that 
a new even IQHE exists
that is distinct from the old one and whose stability relies on the 
existence of the many-particle interactions as well as the occupation of 
higher Landau levels.  In this letter, we investigate the energetic
properties of such a state at Landau level filing fraction $\nu=2$
and argue that it is more stable than the conventional $\nu=2$ state
for small magnetic fields.
 
The investigation of new low magnetic field IQHE states 
is motivated by an experimental phase diagram that is inconsistent 
with an IQHE insensitive to
many-particle interactions \cite{Jiang1,Jiang2,Pepper,Tsui}.
At low magnetic fields, a direct
second order phase transition from a state with 
$\nu=2$ to an insulator ($\nu=0$) has been observed with
no experimental evidence for an intermediate $\nu=1$ phase
in the transition region.
However, in the non-interacting picture, $\nu$ measures the
number of filled Landau levels below the Fermi energy and the
system must decrease the number of these levels by one at a time and
therefore pass through a $\nu=1$ phase in the transition from
$\nu=2$ to $\nu=0$ \cite{Girvin}.
It was therefore proposed that an unconventional $\nu=2$ quantum Hall state,
which can make a direct continuous   
transition to an insulator, arises in low magnetic fields.
This state will be called the $2b$ state and is
distinct from the conventional spin-unresolved
$2a$ state.  
At any finite magnetic field, the $2a$ state is adiabatically connected
to the $2a$ state at infinite magnetic in which the lowest Landau level
for each spin is completely filled.

The essential idea behind the $2b$ quantum Hall state
is that all up- and down-spin electrons pair to form 
spinless bosons.
At electronic filling fraction $\nu=2$, these
bosons form their own Landau levels with filling fraction
$\nu_b=\nu/4=1/2$.  When $\nu_b^{-1}$ is an even integer,
bosons can form a Laughlin state 
that is analogous to the Laughlin fractional quantum Hall
states that exist when $\nu^{-1}$ 
is an odd integer \cite{Girvin}.
The transition from $\nu=2$ to an insulator is the allowed
transition from the $\nu_b=1/2$ bosonic quantum Hall state
to an insulating state.

A proposed spin-singlet many-electron wavefunction
\cite{tikof,Herbut,Permanent,Haldane} 
for the $2b$ state is 
\begin{equation}
\label{wfn}
\Phi=\prod_{k<l,\sigma}(u^\sigma_k v^\sigma_l -
u^\sigma_l v^\sigma_k)\
{\rm Per} \left[F(|u_i v_j - u_j v_i|) \right]\ ,
\end{equation}
where $(u_i,v_i)$ is the complex coordinate of 
the $i^{\rm th}$ electron on the sphere,
$(u^\sigma_k,v^\sigma_k)$ is the complex coordinate of the 
$k^{\rm th}$ spin $\sigma$
electron, and Per denotes the 
Permanent of the symmetric matrix whose $(i,j)^{\rm th}$
component is 
$F(|u_i v_j - u_j v_i|)$.
Because $F$ in not analytic, the $2b$ state has a significant occupation of
higher Landau levels.  This is required because any 
candidate $2b$ state
must be thermodynamically distinct from the conventional
spin-unpolarized $2a$
state, the unique state at $\nu=2$ in the lowest Landau level.
In addition, $\Phi$ has total filling fraction $\nu=2$ as the total
angular momentum of this state is identical to that of the $2a$ state.

The behavior of the $2b$ state described by Eq. (\ref{wfn}) is governed
by the function $F$.  For $F=1$, $\Phi$ is the conventional $2a$ state
consisting of two filled Landau levels.
If $F$ is short-ranged then it can be thought of as a pair wavefunction
and its effective size is the coherence length.  For distances much
longer than this coherence length,
$F$ acts like a $\delta$-function and 
\begin{equation}
\label{wfn2}
\Phi\simeq\prod_{k<l}(u^\uparrow_k v^\uparrow_l -
u^\uparrow_l v^\uparrow_k)^2
{\rm Det} \left[\delta^2
(|u^\uparrow_i v^\downarrow_j - u^\uparrow_j v^\downarrow_i|) \right]\ .
\end{equation}
When $F$ is approximated as a $\delta$-function, 
there is a well-defined pair coordinate and
$\Phi$ vanishes as the second power
of the complex pair coordinate as two pairs approach each other.
$\Phi$ therefore acts like a $\nu_b=1/2$ Laughlin state of charge $e^*=2e$
bosonic pairs.

At first, the existence of a pairing state
stabilized by repulsive Coulomb interactions seems counterintuitive.
Moreover, because a pairing wavefunction is constructed by
occupying higher
Landau levels, we lose kinetic energy that must
be compensated by a gain in interaction energy.
However, we only lose interaction energy among the two
electrons in a pair; we gain repulsive energy 
because the Laughlin $\nu_b=1/2$ state keeps the pairs 
well separated.
The maximum possible 
gain in energy can be estimated by approximating
the pairs as tightly bound point objects thus ignoring the energy cost
for forming the pairs.
We can then calculate the interaction energy for all
electrons not in the same pair.
This is just the Coulomb energy of a $\nu_b=1/2$ state of spinless
charge $e^*=2e$ bosons.  
Laughlin's interpolation formula
for $E_m$ \cite{Girvin},
the Coulomb energy of a projected $\nu=1/m$ state,
gives an energy per electron for the $2b$ state of
$U_{2b}=-0.49 (e^*)^2/\epsilon l^*=-1.39 e^2/\epsilon l$ where $l$ is
the magnetic length.
In contrast, the energy of the $2a$ state is
$E_{2a}=-\sqrt{\pi/8}\ e^2/\epsilon l$.
Forming a paired state has yielded an energy gain per pair of
$2(E_{2a}-U_{2b})=1.53 e^2/\epsilon l$. 

In this paper, we will argue that the gain in interaction energy
between pairs, associated with a $2b$ state,
outweighs the energy cost of forming pairs.
In order to make the formation of pairs as inexpensive
as possible, we consider pairing functions $F$ costing the least
Coulomb energy.  If $F(r)$ is strongly peaked
at $r=0$, electrons in a pair have large overlap 
and pair formation is expensive.
We give $F$ the variational freedom to reduce the cost of
pair formation by taking
\begin{equation}
F(r)=(r-\beta) \exp(-\alpha r)\ , 
\end{equation}
where pairing is indicated by a nonzero inverse pair size $\alpha$
and $\beta$ is a variational parameter
chosen to minimize the cost of pair formation.
If we had only two electrons, one of each spin, then the many body
wavefunction in Eq. (\ref{wfn})
would simply be $F$.  The cost of pair formation, which is
the expectation value of the Coulomb interaction in this state, is 
easily shown to be minimized for $\beta\alpha = 1 -1/\sqrt{2}$.
In contrast, one might have expected that giving 
$F$ a node at $r=0$ and hence taking $\beta=0$ would yield the 
smallest Coulomb energy.  
If we rewrite $F$ as the sum of two terms as
$F(r)=r \exp(-\alpha r) + (-\beta)\exp(-\alpha r)$,
it is easily shown that the energetic gain of having $\beta>0$
comes from the negative
overlap of these two terms.  However, a large
value of the second term still implies a large value of $|F(0)|$ which
costs repulsive Coulomb energy.
We therefore impose the additional restriction that $F(r)$ 
be replaced by $G(r)={\rm Max}(F(r),F(r_o))$ where
$r_o$ depends on $\alpha,\beta$ and is chosen so that not more
$5\%$ of the integral $\int d^2r |G(r)|^2$ comes from $r<\beta$.  
If we fix $\beta$ then $r_o=0$ 
for small $\alpha$, becomes nonzero only for large enough
$\alpha$, and then increases with $\alpha$.

We employ a spherical geometry to calculate the energy of 
$\Phi$ in Eq. (\ref{wfn}) for a finite number of electrons.
Because this choice enforces a uniform density, we avoid the influence
of a physical edge in our calculations \cite{Haldane}.
As is the convention, we
induce a uniform magnetic field by placing a monopole at the center
of the sphere such that the the total magnetic flux through the sphere is
$2S$ flux quanta $hc/e$ and $2S$ is required
to be an integer by the Dirac quantization condition.
The single particle kinetic energy operator on the sphere is 
the conventional $K = |(-i\hbar \nabla +{e\over c}\vec{A})|^2/2m^*$,
defined in spherical coordinates. 
A spherical system with $2N$ electrons occupying $\Phi$, a state with
angular momentum corresponding to 
filling the lowest Landau level of each spin, has $2S=N-1$.
The corresponding magnetic length is
$l_N=\sqrt{\hbar/m^*\omega}=D/\sqrt{2(N-1)}$ where $D$ is the diameter
of the sphere.
In addition, we define the Coulomb interaction between two
electrons $i$ and $j$ to be $V_{ij}=e^2/\epsilon |r_{ij}|$ where
$r_{ij} = u_i v_j - u_j v_i$ is the complex chord distance.

Calculating the expectation value of the energy using $\Phi$ requires
evaluating multi-dimensional integrals.
Integrals of such large
dimension were computed using
a straightforward Monte-Carlo integration on a parallel Cray
T3D supercomputer.  Because the values of the integrand at different
random points are completely independent, the algorithm trivially
parallelized and showed perfect linear speed-up. 
The most time-consuming step was the computation of
the permanents in Eq.~(\ref{wfn}) for which an effective algorithm can
be found in Ref.~\cite{Permanent}.

\begin{figure}[t]
\epsfxsize=8.5 cm
\epsfbox{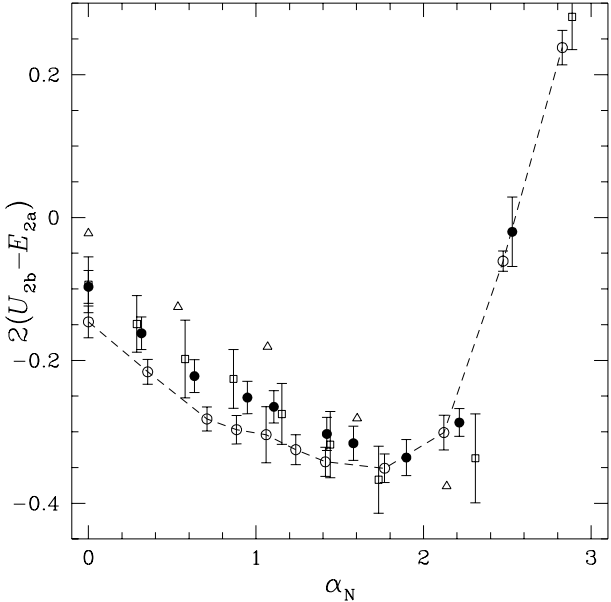}
\caption{The potential energy per pair $2(U_{2b}-E_{2a})$, defined
in units of $e^2/\epsilon l_N$, as a function of $\alpha_N=\alpha l_N$.
Zero is chosen as the energy for the state formed by filling
the lowest Landau for each spin
$2(E_{2a})$ and the data for 2N electrons
are indicated by empty circles $(2N=10)$,
solid circles $(2N=12)$, empty squares $(2N=14)$, and empty triangles
$(2N=16)$.  The dotted line was drawn through the $2N=10$ data as a 
guide for the eye.  The Monte Carlo error bars
for the $2N=16$ data are so large that they were left out for clarity.}
\label{pot-energy}
\end{figure}

The calculated expectation value of the total 
Coulomb energy shows that that the energy gain 
due to the pairing correlations outweighs the energy cost of
forming the pairs.  
The $2b$ state's Coulomb energy is compared to the 
energy of two filled Landau levels, 
$2E_{2a}=-\sqrt{\pi/2}\ e^2/\epsilon l$, where $l$ is the $N\rightarrow\infty$
magnetic length.
In Fig. \ref{pot-energy}, 
the potential energy gain per pair is shown for various system sizes
as a function of $\alpha_N = \alpha l_N$.  For each value of $N$,
$\beta$ is fixed to be that value which gives lowest
expectation value for the total interaction energy at the optimal value
of $\alpha_N$.
We see that the value of $\alpha_N$ at which the energy is minimal
does not change as the size of the system increases.  Therefore, the
gain in energy associated with pairing seems to survive
the thermodynamic limit as 
the preferred value of the pair size $1/\alpha$
scales with the magnetic length $l_N$ and not the system size $D$. 
Because the filling fraction is fixed at $\nu=2$, 
the pair size also scales as the interparticle spacing.
In contrast, the preferred value of $\beta$ seems to be independent
of $N$ and ranges from 0.12$D$ to 0.22$D$.

In addition to the tendency towards pairing,
Fig. \ref{pot-energy} provides further information 
about the behavior of the interaction energy.
Because $F$ is not constant, there is an
energy gain near $\alpha=0$ due to mixing in of higher Landau levels.
When $\alpha D <<1$, the pair size is much greater than the
system size and the energy can not differentiate between different
values of $N$.  In this region, the
interaction energy per particle,
as a function of $\alpha$, should be independent of $N$.  
While this behavior is consistent with our
results, it is not apparent in Fig. \ref{pot-energy}
as the energy is plotted as
a function of $\alpha_N$ instead of $\alpha$ in addition to being
defined in units of $e^2/\epsilon l_N$ instead of $e^2/\epsilon D$.
The energy is linear in the inverse pair size $\alpha_N$ in the 
$\alpha\rightarrow\infty$ limit as it is 
dominated by the Coulombic 
energy cost of forming pairs.

\input epsf

\begin{figure}[t]
\epsfxsize=8.5 cm
\epsfbox{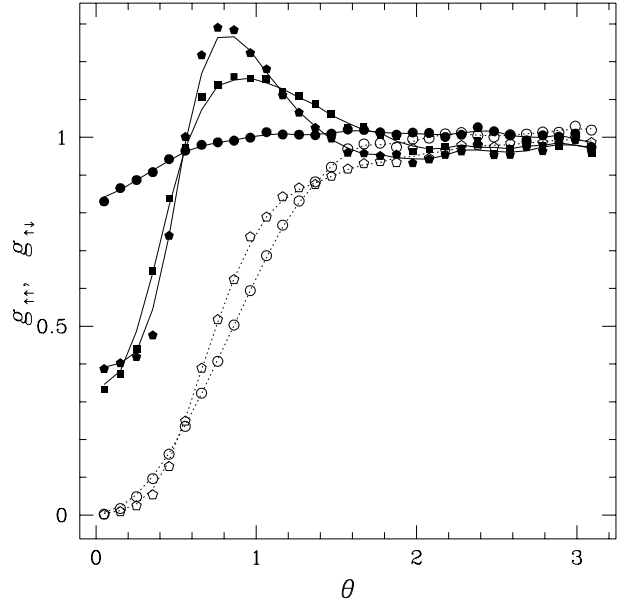}
\caption{The correlation functions are shown for $2N=10$ electrons
as a function of the angle $\theta$ where $D\sin{\theta/2}$ is the
chord distance.
Values of $g_{\uparrow\downarrow}$ and $g_{\uparrow\uparrow}$
are denoted by the solid markers and empty markers, respectively.
The $\alpha=0$ data are indicated by circles; the data for
$\alpha=5$, the optimal value of $\alpha$, are indicated by pentagons;
and the data for $\alpha=3$ data are indicated by solid squares.
The lines are only meant as guides to the eye.
As is the convention, the correlation functions are normalized so
that at large distances (in the $D\rightarrow\infty$ limit)
$g_{\sigma,\sigma'}(\pi)=1$.
}
\label{correlations}
\end{figure}

In Fig. \ref{correlations},
pairing is shown to affect the correlations 
between the locations of electrons. 
It is clear from the expression for the energy per pair
of a system with spin-rotation symmetry and electron density $\rho$,
\begin{equation}
U_{2b}={\rho\over 2}\int d^2r \left(g_{\uparrow\uparrow}(r)
+g_{\uparrow\downarrow}(r)-2 \right){e^2\over \epsilon r}\ ,
\end{equation}
that modifications of the correlation functions are
responsible for the energy gain shown in Fig. \ref{pot-energy}.
There is only a minor change in the behavior of the same spin correlation 
function $g_{\uparrow\uparrow}$ as we vary from $\alpha$ from $0$ to
its energetically preferred value.
The $2a$ state does so well at keeping same
spin particles apart that the $2b$ state need not do much better.
On the other hand, 
simply filling two Landau levels yields no correlations
between up and down spins and a constant $g_{\uparrow\downarrow}$.
The $2a$ state, given by $\Phi$ in Eq. (\ref{wfn}) with $\alpha=0$, 
has only weak correlations between opposite spin electrons.
Therefore, having a finite $\alpha$ is favored
energetically because it introduces strong correlations between
the locations of up and down spins.  These correlations minimize the
repulsive Coulomb energy between up and down spins by giving them
a preferred separation and thus keeping them on average farther apart.

\begin{figure}[t]
\epsfxsize=8.5 cm
\epsfbox{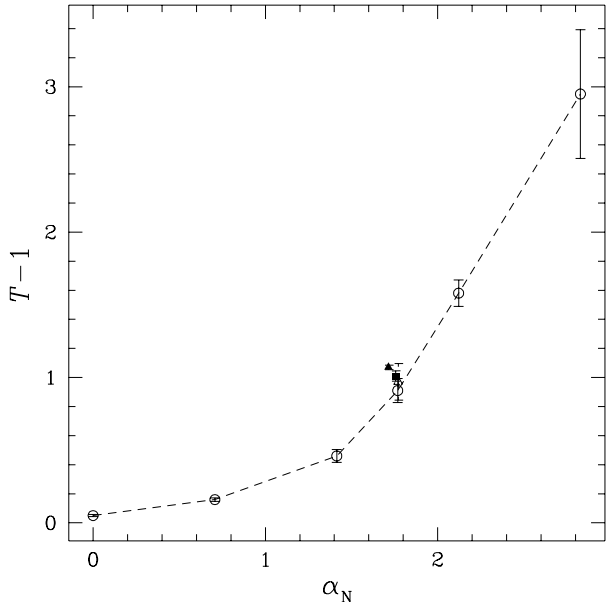}
\caption{The kinetic energy per pair $T$,
defined in units of the
cyclotron energy $\hbar \omega_c$.  Zero is chosen to be the kinetic
energy of the $2a$ state, $\hbar\omega_c$.
The solid circles are data
for $2N=10$ electrons.  Convergence to the thermodynamic
limit is shown at one value of $\alpha_N$ and the data
for $2N=6$ (solid square), $2N=8$ (solid triangle), and $2N=12$
(open triangle) are approximately within each other's error bars.
Because the dominant length scale determining the kinetic energy
in the large $\alpha$ limit is the pair size $1/\alpha>>l$, the
kinetic energy scales as $\alpha^2$ for large $\alpha$.}
\label{kinetic-energy}
\end{figure}

In spite of mixing in all higher Landau levels,
the loss of kinetic energy associated with the $2b$ state is only of order
the cyclotron energy $\hbar \omega_c$ per pair as is seen in 
Fig. \ref{kinetic-energy}.
The prefactor $\prod_{i<j,\sigma} (u_i^\sigma v_j^\sigma
-u_j^\sigma v_i^\sigma)$
forces much of the $2b$ wavefunction into
the lowest Landau level.  The ability of a Jastrow prefactor
to force most of the wavefunction into the lowest Landau level is
well known from Jain's construction of hierarchical quantum Hall
states \cite{Jain}.
However, that work assumed that the wavefunction was not thermodynamically
distinct from a wavefunction projected onto the lowest Landau level.
In contrast, a paired wavefunction must necessarily involve all Landau
levels.  The two-particle paired wavefunction,
$\Phi\sim F(r)$, can be expressed in terms of its
projection onto the $n^{\rm th}$ Landau level as $\Phi=\sum_n a_n \psi_n(r)$
where $\psi_n(r)$ is the zero angular momentum state in the $n^{\rm th}$ 
Landau level.  If we do not include all terms in this series and hence
all Landau levels then we do not really have pairing as $F(r)$ will 
oscillate instead of vanishing exponentially at large $r$.

The energy gain associated with the $2b$ state can outweigh the energy loss
only for small enough magnetic field. 
The energy gain is Coulombic and scales as $e^2/\epsilon l \sim \sqrt{B}$
while the energy loss is kinetic and scales as $\hbar \omega_c \sim {B}$.
Therefore, there is an energy gain associated with forming a
$2b$ state only for
large enough values of the ratio
\begin{equation}
y={e^2/\epsilon l\over  \hbar \omega_c}=l/a^*=\sqrt{B^*\over B}\ ,
\end{equation}
where $a^*=\epsilon \hbar^2 /m^*$ is
the effective Bohr radius.
For GaAs$-$Al$_x$Ga$_{1-x}$As, $a^*=100 \rm \AA$ and $B^*=6.6 T$.
Our calculation first gives a gain in energy for $y\sim 1$.

The numerical evidence presented does not prove unequivocally that the $2b$
state is the preferred state.
A definitive calculation would compare
the lowest energy $2b$ state to the lowest energy $2a$ state.
Clearly, the specific form of these states would depend on the
value of the external magnetic field $B$ in a complicated way.
In addition, even though the lowest energy $2a$ and $2b$ states are
thermodynamically distinct, they need not
differ greatly in energy. 
It is clearly not feasible to perform an unequivocal energetic comparison
of the $2a$ and $2b$ states.
Instead, we 
compared our $2b$ state to the $2a$ state without any occupation of higher
Landau levels.  In so doing, we have
made the case that
the energetic gain of pair correlations can outweigh the energetic
cost of forming pairs.  In order to do any better, we need to include
the effect of disorder as this is known to stabilize the experimental
signatures of the $2b$ state at larger magnetic 
fields \cite{Jiang1,Pepper}.

The energetic arguments used to justify pairing can be extended
to other quantum Hall systems.  For instance, the same
mechanism that we have discussed would encourage the formation
of a state at $\nu=m$, consisting
of charge $e^*=me$ particles forming a quantum Hall state at composite
filling fraction $\nu_{comp}=1/m$, that makes a direct continuous
transition to an insulator.
In fact, there is recent evidence for a $\nu=3$
to insulator transition \cite{Song}.
In addition, we are presently investigating a
class of paired states at $\nu=1/2$
in double layer systems that are distinct from the
states that have been proposed to explain the experimental
observation of a quantum Hall effect at $\nu=1/2$ \cite{bilayer}.
A preliminary indication of this paired state would be a direct
continuous transition from
the $\nu=1/2$ state to an insulator.

We are grateful to H. W. Jiang, D. Morse, and Y. Yakar for many useful
conversations.  We are especially grateful to S. A. Kivelson for
sharing his coffee and insights with us.
This work has been supported by the National Science Foundation
under grant PHY94-07194 at the Institute for Theoretical Physics
and grant DMR93-08011 at UCSB.  F. G. P. acknowledges the support of the
Center for Quantized Electronic Structures of UCSB and the
Quantum Institute of UCSB.  We also acknowledge a generous grant
of computer time by the San Diego Supercomputer Center where the
calculations were performed.

\end{document}